\documentclass[12pt]{article}
\usepackage{amsthm,graphics}
\newtheorem{theorem}{Theorem}[section]
\newtheorem{definition}[theorem]{Definition}
\newtheorem{lemma}[theorem]{Lemma}
\newtheorem{corollary}[theorem]{Corollary}
\newenvironment{keywords}{\begin{quote}{\bf Keywords:}}{\end{quote}}
\newtheorem{proposition}[theorem]{Proposition}
\usepackage{hyperref}

\usepackage[utf8]{inputenc}
\usepackage[english]{babel}
\usepackage{amsmath,amssymb}
\newtheorem{example}[theorem]{Example}
\newtheorem{remark}[theorem]{Remark}
\usepackage{enumerate}
\usepackage{color}
\usepackage{cancel}

\newenvironment{msccodes}{\begin{quote}{\bf MSC classification codes:}}{\end{quote}}

\def\fq{\mathbb{F}_{q}}
\def\fqs{\mathbb{F}_{q^2}}
\def\fqn{\mathbb{F}_{q^n}}
\def\Pla{\mathcal{P}}
\def\cF{\mathcal{F}}
\def\cA{\mathcal{A}}
\def\cX{\mathcal{X}}
\def\cC{\mathcal{C}}
\def\cF{\mathcal{F}}
\def\cP{\mathcal{P}}
\def\cG{\mathcal{G}}
\def\cH{\mathcal{H}}
\def\cL{\mathcal{L}}
\def\cS{\mathcal{S}}
\def\N{\mathbb{N}}
\def\Z{\mathbb{Z}}
\def\cD{\mathcal{D}}
\def\cF{\mathcal{F}}
\def\a{\alpha}
\def\bb{{\boldsymbol\beta}}
\def\b{\beta}
\DeclareMathOperator\supp{supp}

\newcommand\lu[1]{{\color{red} #1}}
\newcommand\al[1]{{\color{blue} #1}}
\newcommand\ma[1]{{\color{green} #1}}

 \author{Maria Bras-Amorós\thanks{Maria Bras-Amorós is with Universitat Rovira i Virgili, Av. Pa\"isos Catalans 26, 43007 Tarragona, Catalonia, Spain, {\em email: maria.bras@urv.cat} } \, Alonso S. Castellanos\thanks{Alonso S. Castellanos is with Faculdade de Matem\'atica - Bloco 1F, Universidade Federal de Uberl\^andia - Av. João Naves de \'Avila, 2121 - B. Santa M\^onica, CEP 38.408-100   - Brazil, {\em email: } alonso.castellanos@ufu.br }, 
and Luciane Quoos\thanks{Luciane Quoos is with Universidade Federal do Rio de Janeiro, Centro de Tecnologia, Cidade Universitária - Av. Athos da Silveira Ramos, 149 - Ilha do Fundão, CEP 21.941-909 - Brazil, {\em email: } luciane@im.ufrj.br}
}

\title{Isometry-Dual Flags of Many-Point AG Codes}

\date{\today}

\begin{document}

\maketitle

\enlargethispage{2cm}

\begin{abstract}
 A flag of linear codes $C_0 \subsetneq C_1 \subsetneq \cdots \subsetneq C_s$ is said to have the {\it isometry-dual property}  if there exists a vector ${\bf x}\in (\mathbb{F}_q^*)^n$ such that  $C_i={\bf x} \cdot C_{s-i}^\perp$, where $C_i^\perp$ denotes the dual code of the code $C_i$. We extend our previous results in \cite{BCQ2021}, of flags of algebraic geometry two-point codes over a function field $\cF$ to flags of $(t+1)$-point codes 
$$C_\mathcal L(D, a_0P+\sum_{i=1}^t\beta_iQ_i)\subsetneq C_\mathcal L(D, a_1P+\sum_{i=1}^t\beta_iQ_i))\subsetneq \dots \subsetneq C_\mathcal L(D, a_sP+\sum_{i=1}^t\beta_iQ_i)$$  
  for {\em any} tuple of integers $\beta_1,\dots,\beta_t$ and for an increasing sequence of integers $a_0,\dots,a_s$, just provided that $n\geq 2g+2$, where $g$ is the genus of $\cF$.
  We apply the obtained results to the broad class of Kummer extensions defined by affine equations of the form $y^m=f(x)$, for $f(x)$ a separable polynomial of degree $r$, where $\gcd(r, m)=1$. 
  In particular, we obtain necessary and sufficient conditions on $m$ and $\beta_i$'s such that the flag has the isometry-dual property.
\end{abstract}


\begin{msccodes}
  14G50,        
  11T71,   	
  94B27,   	
  14Q05   	
\end{msccodes}

\section{Introduction}
Let $\fq$ be the finite field with $q$ elements. A linear code $C$ of $\fq$ is a $\fq$-linear subspace of $\fq^n, n\geq 1$.  
Algebraic geometry codes (abbreviated by AG codes) constitute a special family of linear codes that makes use of algebraic function fields in its construction. 
Already in 1982,
Tsfasman, Vl\c{a}dut and Zink \cite{TVZ1982} put the focus to these codes when they proved the existence of AG codes with surprising asymptotic performance exceeding the well-know Gilbert-Varshamov bound.
In 1995, Garcia and Stichtenoth \cite{GS1995} constructed explicit sequences of AG codes providing a complete and self-contained proof of the Tsfasman-Vladut-Zink theorem. This result is a beautiful and important application of function fields to coding theory.
Nowadays AG codes are actively used in a variety of research fields such as quantum codes \cite{LP2017, MTF2016}, in locally recoverable codes \cite{BMQ2020}, in secret sharing \cite{M2018} and in cryptography \cite{CMP2017}.

Let  $\cF/\fq$ be a function field. Consider $D=P_1+\dots+P_n$ a divisor given by the sum of pairwise distinct rational places of $\cF$,  and $G$ a divisor such that $P_i$ is not in the support of $G$ for $i = 1,\dots ,n$. The algebraic geometry code $C_{\mathcal{L}}(D, G)$  is defined as
\begin{equation}C_{\mathcal{L}}(D, G)=\{ (f(P_1),\ldots, f(P_n))\, :\, f\in \mathcal{L}(G)\} \subseteq \fq^n,\label{defAGcode}\end{equation}
where $\cL(G)=\{z\in \cF\,:\, (z)\ge -G\}\cup \{0\}$ denotes the Riemann-Roch space associated to the divisor $G$. 
These codes are called {\it one-point} or {\it many-point codes} provided that the divisor $G$ is supported on a single rational place $P$, that is, $G=aP$, or in many pairwise distinct rational places $Q_1, \dots , Q_t$, that is, $G=\b_1Q_1+ \cdots + \b_tQ_t$, respectively, where $a, \b_1,\dots, \b_t \in {\mathbb Z}$.
Determining or even improving the parameters of AG codes has been a major subject of research, see for example \cite{B2009,  BO2008, CT2005, DK2011,GMRT2011}, and \cite{JX2018}.

The dual code $C^\perp$ of a linear code $C$ is the orthogonal complement of $C$ in $\fq^n$ with the standard inner product of $\fq^n$. The knowledge of dual codes is specially important in the detection and correction of errors.
For an AG code $C_{\mathcal{L}}(D, G)$, its dual is a $C_{\mathcal{L}}(D, H)$ code for a suitable divisor $H$ depending on a Weil differential with certain properties that make it not easy to be explicitly determined, see \cite[Proposition 2.2.10]{St2009}. 

A code is said to be {\em self-dual} if $C=C^\perp$.
Self-dual codes have been investigated in 
 \cite{MST2008} and have applications to quantum codes through a construction in \cite{KM2008}, see for example \cite{LP2017, MTF2016}. 
 
Different names have been given in the literature to pairs of linear codes $C_1, C_2$ such that there exists ${\bf x}\in (\mathbb{F}_q^*)^n$ such that $C_1={\bf x}\cdot C_2$. They are said to be either equivalent, for instance in \cite[Definition 2.2.13]{St2009}, or isometric, for instance in \cite[Subsection 4.1]{GMRT2011}.
 
The restrictive condition of self-dual codes can be relaxed 
to codes that are equivalent to their duals, or, even further, to flags of codes 
$C_0 \subsetneq C_1 \subsetneq \cdots \subsetneq C_s$ 
such that 
$C_i$ is {\bf x}-isometric to $C_{s-i}^\bot$ for all $i=0,\ldots, s$, for a unique constant vector ${\bf x}\in (\mathbb{F}_q^*)^n$. These flags of codes are said to satisfy the {\it isometry-dual property}. In the case of flags of AG codes, having the isometry-dual property gives directly dual codes without the need of computing Weil differentials.

Natural increasing flags of one-point codes are obtained by varying $a\geq 0$ in the divisor $G = aP$, in $C_{\mathcal{L}}(D, G)$. Geil, Munuera, Ruano and  Torres \cite{GMRT2011} first introduced the notion of the isometry-dual property for analyzing the behavior of the minimum distance (and so the error-correction capability) along these flags of one-point AG codes. 
In \cite{BDH2020} there is an analysis of the effect of puncturing isometry-dual flags of codes.
The analysis of the minimum distance along a sequence was then extended, although using different words, by Kim and Lee \cite{BY2018} for two point codes. In particular, the isometry-dual property was studied for two-point codes in \cite{BCQ2021}. Then, the same notion of isometry-dual flags of codes was used by Munuera, Tenorio, and Torres to construct quantum codes.

In reference \cite{BCQ2021} we analyzed flags of two-point codes $C_{\mathcal{L}}(D, aP+\beta Q)$, obtained by fixing $\beta\in{\mathbb N}$ and varying $a \geq 0$.
It remained to extend our results to many-point codes. Indeed, 
Matthews proved that many-point codes on a curve may have better parameters than any comparable one-point code over the same curve \cite{G2005},
and  many-point codes appeared as codes with best known minimum distance for fixed length and dimension in \cite{BQZ2020}. 
 Also, while for one-point codes there is only need to analyze positive integers $a$, for the case of two-point codes, as well as for $(t+1)$-point codes in general, the integer $a$ may be negative.  
In this contribution we extend our previous results in two directions. On one hand we extend our results to the case $C_\mathcal L(D, aP+\sum_{i=1}^t\beta_iQ_i)$ of $(t+1)$-point codes as suggested by the aforementioned results and, on the other hand, we extend our results to the case of negative integers $a$ as well as to the case of negative coefficients $\beta_i$. By allowing negative coefficients in the defining divisors we obtain new flags of codes for each given curve (the ones corresponding to negative values of $\beta_i$), and at the same time the flags are larger in the sense that they are formed by a larger number of codes (because of the newly allowed values of $a$).

To consider flags of many-point algebraic geometry codes we deal with the generalized Weierstrass semigroups introduced in \cite{BT2006} and also investigated in \cite{MTT2019, TeT2019}. For $\mathbf{Q}_t=(Q_1,\ldots, Q_t)$ a tuple of pairwise distinct rational places on the function field $\cF$, the  generalized Weierstrass semigroup of $\cF$ at $\mathbf{Q}_t$ is the set
\begin{equation}
\widehat{H}(\mathbf{Q}_t)=\{(-\upsilon_{Q_1}(h),\ldots,-\upsilon_{Q_t}(h))\in\Z^t:h\in R_{\mathbf{Q}_t}\}\,,
\end{equation}
where $\upsilon_{Q_i}$ stands for the valuation of the function field $\cF$ associated to $Q_i$, and $R_{\mathbf{Q}_t}$ stands for the ring of functions of $\cF$ being regular out of the set $\{Q_1,\ldots, Q_t\}$.

Given ${\bb}=(\b_1,\ldots, \b_t)$ a fixed tuple in ${\mathbb Z}^t$ we consider the divisor ${\bf G_{\bb} }=\b_1Q_1+\cdots+\b_tQ_t$ and  investigate the isometry-dual property in flags of many-point algebraic geometry codes, that is,
$$
C_\mathcal L(D, a_0P+{\bf G_\bb})\subsetneq
C_\mathcal L(D, a_1P+{\bf G_\bb})\subsetneq \dots \subsetneq C_\mathcal L(D, a_sP+{\bf G_\bb}).
$$

In order to analyze the dimension jumps in these flags of many-point codes, we first provide results related to the sets
$$
\widehat{H}_{\bf \bb}=\{a\in\Z  \, :\,  \ell(aP+{\bf G_{\bb}})\neq \ell((a-1)P+{\bf G_{\bb}}) \}
$$
and 
$$
\widehat{H}_{\bf \bb}^*=\{a\in\Z \, :\, C_\mathcal L(D, aP+{\bf G_\bb})\neq C_\mathcal L(D, (a-1)P+{\bf G_\bb}) \}\;.
$$
They are closely related to the Weierstrass semigroup at one rational place $H(P)$, and the generalized Weierstrass semigroups in many rational places $\widehat{H}(P, \mathbf{Q}_t)=\widehat{H}(P, Q_1, \ldots, Q_t)$.
Then we prove in Theorem~\ref{t:isodualKummer2} a characterization of the tuples of integers $\beta_1,\dots,\beta_t$ for which the flag of codes satisfies the isometry-dual condition in terms of the set $\widehat{H}_{\bf \bb}^*$, the code length and the genus of the curve.

While our characterization of the integers $\beta_1,\dots,\beta_t$ giving isometry-dual flags of codes applies for general function fields, many of the function fields used in coding theory are described by an affine equation of the form $y^m=f(x)$ for some polynomial $f(x) \in \fq[x]$. It is the case of the following maximal function fields:  Hermitian function field \cite{MMP2005}, the Giullietti-Korchmáros function field \cite{GK2009},  the generalized Hermitian function field described in \cite{KKO2001} and the curves with many-points described in \cite{MRQ2021} and references in there. 
These function fields are appreciated, in particular, because they are related to maximal curves and they guarantee a large code length. 
Function fields defined by an affine equation of the form $y^m=f(x)$ are known as Kummer extensions, see \cite{St2009}.

At the end of our work we focus on Kummer extensions given by $y^m=f(x)$, where $f(x)$ is a separable polynomial of degree $r$, with $r, m$ coprime and $2\leq r\leq m-1$. Using a result by Maharaj \cite{M2004} we give a complete description of the sets $\widehat{H}_{\bf \bb}$ and $\widehat{H}_{\bf \bb}^*$. Then we characterize the tuples $\beta_1,\dots,\beta_t$ for which the related flags of codes satisfy the isometry-dual condition just in terms of divisibility among the integers $\beta_i$ in the tuple and $m$. This is stated in Theorem~\ref{t:isodualKummer3}. We provide examples of flags of many-point codes over the Klein and Hermitian function fields in examples \ref{klein} and \ref{examples} respectively.

The article is organized as follows. In Section 2 we collect some results on function fields and generalized Weierstrass semigroups. In Section 3 we present results on the relationships among the sets $\widehat{H}_{\mathbf {\bb}}, \widehat{H}_{\mathbf {\bb}}^*$, the Weierstrass semigroup  $H(P)$, and the generalized Weierstrass semigroup $\widehat{H}(P, {\mathbf{ Q}}_t)=\widehat{H}(P, Q_1, \ldots, Q_t)$. 

In Section 4 we prove conditions to obtain flags satisfying the isometry-dual property for general algebraic geometry codes on many points over a function field, see Theorem \ref{t:isodual}.
In Section 5 we introduce two kinds of AG codes on many points over Kummer extensions and, for each one of them, we characterize the integers $a$ that belong to $\widehat{H}_{\mathbf {\bb}}^*$ and investigate the isometry-dual property.

\section{Preliminary results}

\subsubsection*{Function fields and algebraic geometry codes}

Let $\cX$ be a projective, absolutely irreducible, nonsingular algebraic curve of genus $g$ defined over the finite field $\fq$. We denote by $\cF/\fq$  its function field with respect to  the field of constants $\fq$. For a function $z\in\cF$, let $(z)$, $(z)_0$ and $(z)_\infty$ stand for its principal, zero and pole divisor, respectively. We denote by $\Pla(\cF)$ the set of places of $\cF$ and by $\cD_{\cF}$ the free abelian group generated by the places of $\cF$. The elements $D$ in $\cD_{\cF}$ are called {\it divisors} and can be written as
$$
D=\sum_{P\in \Pla(\cF)}n_P\,P\quad \text{ with } n_P\in \Z,\;  n_P=0\text{ for almost all }P\in\Pla(\cF).
$$
The degree of a divisor $D$ is the sum $\deg(D)=\sum\limits_{P\in \Pla(\cF)}n_P \cdot\deg P$, where $\deg P$ is the degree of the place $P$ over $\fq$.
Given a divisor $G\in\cD_{\cF}$, the Riemann-Roch vector space associated to $D$ is defined by
$$
\cL(G):=\{z\in \cF\,|\, (z)\ge -G\}\cup \{0\}.
$$
Its dimension with respect to the field of constants $\fq$ is denoted $\ell(G)$. Two divisors $G_1$ and $G_2$ in $\cD_\cF$ are said to be equivalent if there exists a function $z \in \cF$ such that $G_1=G_2 + (z)$ and we write $G_1 \sim G_2$.
In this case $\cL(G_1)=z^{-1}\cL(G_2)$ and it follows that the Riemann–Roch spaces $\cL(G_1)$ and  $\cL(G_2)$ are isomorphic.

For  $C_\cL(D,G)$ an AG code, through all the paper $n$, $k$, $d$ stand for the length, dimension, and minimum distance of the code, respectively. It holds
  \begin{equation}\label{dimcodedivisordifference}k=\ell(G)-\ell(G-D)\end{equation}
  and, 
by the Riemann-Roch theorem \cite[Th. 1.5.17]{St2009},
  \begin{equation}\label{dimcode}
 \begin{split}
&\text{ if } 2g-2<\deg(G) < n, \text{ then } k=\ell(G)= \deg(G)+1-g.
\end{split}
\end{equation}
An interesting and well known property about AG codes is that the dual of $C_{\mathcal{L}}(D, G)$ is still an $C_{\mathcal{L}}(D, H)$ code for a suitable divisor $H$. In fact, from \cite[Proposition 2.2.10]{St2009}, we have that 
\begin{equation}\label{dualcode}
C_{\mathcal{L}}(D, G)^\perp=C_{\mathcal{L}}(D, D-G+(\eta)),
\end{equation}
where $\eta$ is a Weil differential such that $v_{P_i}(\eta)=-1$ and $\eta_{P_i}(1)=1$ for any $P_i$ in $\supp(D)$.

\subsubsection*{Generalized Weierstrass semigroups}

Let $\mathbb{N}=\{0, 1, \dots \}$ denote the set of natural numbers and suppose the function field $\cF/\fq$ has genus $g\geq 1$. Given an $t$-tuple $\mathbf{Q}_t=(Q_1,\ldots, Q_t)$ of $t$ pairwise distinct rational places $Q_i$ on $\Pla(\cF)$, the {\em Weierstrass semigroup} in the $t$ places is
$$
H(\mathbf{Q}_t)= \{(\b_1, \dots , \b_t) \in \N^t \ : \ \exists \ z \in \cF \text{ with } (z)_\infty = \b_1Q_1 + \cdots + \b_tQ_t \}.
$$
The complement $G(\mathbf{Q}_t)=\N^t \setminus H(\mathbf{Q}_t)$ is always a finite set and its elements are called {\it Weierstrass gaps} at $Q_1, \dots, Q_t$. A gap can be characterized in terms of the dimension of certain Riemann-Roch spaces, in fact, a $t$-tuple $(\a_1, \dots , \a_t) \in \N^t$ is a gap at $Q_1, \dots, Q_t$ if and only if 
$$\ell\left(\sum_{i=1}^t \a_iQ_i\right)= \ell\left((\sum_{i=1}^t \a_iQ_i)- Q_j\right) \text{ for some } j \in \{1, \dots, t\}.$$
 
Beelen and Tuta\c s generalized the notion of a Weierstrass semigroup by allowing the $t$-tuples to assume values on the integers. We collect some results that can be found in \cite{BT2006}.
\begin{definition}
 For $\mathbf{Q}_t=(Q_1,\ldots, Q_t)$ a tuple of pairwise distinct rational places on the function field $\cF$, the  generalized Weierstrass semigroup of $\cF$ at $\mathbf{Q}_t$ is the set
\begin{equation}
\widehat{H}(\mathbf{Q}_t)=\{(-\upsilon_{Q_1}(h),\ldots,-\upsilon_{Q_t}(h))\in\Z^t:h\in R_{\mathbf{Q}_t}\}\,,
\end{equation}
where $\upsilon_{Q_i}$ stands for the valuation of the function field $\cF$ associated to $Q_i$, and $R_{\mathbf{Q}_t}$ stands for the ring of functions of $\cF$ being regular out of the set $\{Q_1,\ldots, Q_t\}$.
\end{definition}
Provided $q\geq t$, the classical and generalized Weierstrass semigroups are related by 
 $$H({\mathbf Q}_t)=\widehat{H}({\mathbf Q}_t)\cap \N^t.$$
It is clear that if $(\a_1,\ldots,\a_t)\in \widehat{H}({\bf Q}_t)$, then $\a_1+\cdots+\a_t\geq 0$, and if $\a_1+\cdots+\a_t\geq 2g$, then $(\a_1,\ldots,\a_t)\in\widehat{H}({\bf Q}_t)$.
\begin{lemma}
Given a $t$-tuple $\mathbf{Q}_t=(Q_1,\ldots, Q_t)$ of $t$ pairwise distinct rational places on the function field $\cF/\fq$, 
for any $s=1, \ldots, t-1$ we have $\widehat{H}(Q_1,\ldots,Q_s)\times\widehat{H}(Q_{s+1},\ldots,Q_t)\subseteq \widehat{H}({\bf Q}_t)$.
\end{lemma}
\begin{proof}
Let $(\a_1,\ldots, \a_s)\in \widehat{H}(Q_1,\ldots, Q_s)$ and $(\a_{s+1},\ldots, \a_t)\in \widehat{H}(Q_{s+1},\ldots, Q_t)$. Then there exist functions $h \in R_{(Q_1,\ldots, Q_s)}$ and $g \in R_{(Q_{s+1},\ldots, Q_t)}$  such that $(\a_1,\ldots, \a_s)=(-\upsilon_{Q_1}(h),\ldots,-\upsilon_{Q_s}(h))$ and $(\a_{s+1},\ldots, \a_t)=(-\upsilon_{Q_{s+1}}(g),\ldots,-\upsilon_{Q_t}(g))$. By definition, we have that $\upsilon_{Q_i}(h)\geq 0$ for $i=s+1,\ldots, t$ and $\upsilon_{Q_i}(g)\geq 0$ for $i=1,\ldots, s$.  Since $q\geq t$, we can choose $a, a'\in\mathbb{F}_q$ such that $\upsilon_{Q_i}(h+a)=0$ for all $i=s+1,\ldots, t$ and $\upsilon_{Q_i}(g+a')=0$ for all $i=1,\ldots, s$. So, $(\a_1,\ldots, \a_s,0,\ldots, 0),(0,\ldots,0, \a_{s+1},\ldots,\a_t)\in\widehat{H}({\bf Q}_t)$ and therefore $(\a_1,\ldots, \a_t)\in\widehat{H}({\bf Q}_t)$.
\end{proof}

For ${\mathbf\alpha}=(\alpha_1,\ldots,\alpha_t)\in\Z^t$, $G_{\mathbf\alpha}$ will denote the divisor $\alpha_1 Q_1+\cdots+\alpha_tQ_t$ on $\cX$. Given $i\in\{1,\ldots, t\}$, let 
$$
\nabla_i^t({\mathbf\alpha}):=\{{\mathbf \beta}=(\beta_1,\ldots,\beta_t)\in \widehat{H}({\mathbf Q}_t):\beta_i=\alpha_i\mbox{ and } \beta_j\leq\alpha_j\mbox{ for }j\neq i\}\;.
$$

\begin{proposition}\cite[Proposition 2.1]{MTT2019}\label{Propositiondimension}
Let ${\mathbf\alpha}\in \Z^t$ and assume that $q\geq t$. Then,
\begin{enumerate}[(i)]
\item ${\mathbf\alpha}\in\widehat{H}({\mathbf Q}_t)$ if and only if $\ell(G_{\mathbf\alpha})=\ell(G_{\mathbf\alpha}-Q_i)+1$ for all $i\in \{1,\ldots,t\}$;
	
\item $\nabla_i^t({\mathbf\alpha})=\emptyset$ if and only if $\ell(G_{\mathbf\alpha})=\ell(G_{\mathbf\alpha}-Q_i)$.
\end{enumerate}
\end{proposition}

In \cite{MTT2019} a generating set $\widehat{\Gamma}({\mathbf Q}_t) \subseteq \widehat{H}({\mathbf Q}_t)$ is constructed, which generates $\widehat{H}({\mathbf Q}_t)$ by means of the so-called {\it least upper bound} (lub) of a finite subset $\mathcal{B}\subseteq \Z^t$, that is,
$$
  \mbox{lub}(\mathcal{B}):=(\max\{\beta_1:{\mathbf\beta}\in \mathcal{B}\},\ldots,\max\{\beta_t:{\mathbf\beta}\in\mathcal{B}\})\in \Z^t\;.
$$
Then, 
$$\widehat{H}({\mathbf Q}_t)=\{\mbox{lub}({\mathbf\beta}^1,\ldots,{\mathbf\beta}^t):{\mathbf\beta}^1,\ldots,{\mathbf\beta}^t\in \widehat{\Gamma}({\mathbf Q}_t)\}\;.$$ 

One consequence is that $\widehat{H}({\mathbf Q}_t)$ is closed under the operation $lub$. That is,

\begin{equation}\label{e:lubclosed}
  \{\mbox{lub}(\beta^1,\dots,\beta^t):(\beta^1,\dots,\beta^t)\in \widehat{H}({\mathbf Q}_t)\}\subseteq \widehat{H}({\mathbf Q}_t).
  \end{equation}

The set $\widehat{\Gamma}({\mathbf Q}_t)$ is not necessarily finite.

\subsubsection*{Kummer extensions of the form $y^m=f(x)$}

In the present work, we illustrate our results on the broad class of Kummer extensions defined by affine equations of the form $y^m=f(x)$, for $f(x)$ a separable polynomial of degree $r$ with $\gcd(m, r)=1$ and $2\leq r \leq m-1$.
This class contains important examples of curves such as the Hermitian curve, Geil's norm-trace curve \cite{G2003}, the generalized Hermitian function field defined by the equation $y^{q^\ell+1}=x^q+x$ over $\mathbb{F}_{q^{2\ell}}$, as well as the Giullietti-Korchmáros curve. In this case, the dimension of Riemann-Roch spaces can be analyzed by means of the next theorem due to Maharaj \cite{M2004}
by decomposing them as a direct sum of Riemann–Roch spaces of divisors of the projective line. At first we need a definition. For any function field extension $K \subseteq E\subseteq \cF$ over $\fq$, and for a divisor $D$ of $\cF$, define the restriction of $D=\sum_{P\in\Pla(\cF)}n_PP$ to $E$ as the divisor 
$$D_{|E}= \sum_{R\in \Pla(E)} min\,\bigg\{\bigg\lfloor\frac{n_P}{e(P|R)}\bigg\rfloor:  P\in\Pla(\cF)\mbox{ and }{P|R}\bigg\}\,R,$$ where $e(P|R)$ is the ramification index of $P$ over~$R$.
\begin{theorem}\cite[Theorem 2.2]{M2004}\label{ThMaharaj}
  Let $\cF/K(x)$ be a Kummer extension of degree $m$ defined by $y^m=f(x)$.
  Then for any divisor $D$ of $\cF$,  with $D$ invariant by the action of $Gal(\cF/K(x))$, we have that
$$ \mathcal{L}(D)= \bigoplus\limits_{t=0}^{m-1}  \mathcal{L}([D+(y^t)]_{|K(x)})\,y^t.$$
\end{theorem}

\section{Dimension of many-point Riemann-Roch spaces and codes}

Let $t\geq 1$ and $P, Q_1,\ldots, Q_t$ be $t+1$ pairwise distinct rational places in the function field $\cF/\fq$ of genus $g$. For a $t$-tuple ${\bb}=(\b_1,\ldots, \b_t)$ in ${\mathbb Z}^t$ we fix the following notation
$${\bf Q}_t=(Q_1,\ldots, Q_t), \quad {\bf G_{\bb} }=\b_1Q_1+\cdots+\b_tQ_t$$ 
and the set 
$$\widehat{H}_{\bf \bb}=\{a\in\Z  \, :\,  \ell(aP+{\bf G_{\bb}})\neq \ell((a-1)P+{\bf G_{\bb}}) \}.$$
We notice that, if $a$ in $\widehat{H}_{\bf \bb}$, then $a+(\b_1+\dots +\b_t)\geq 0$. For any $a \in \Z$ and a tuple ${\bf \bb}=(\b_1,\ldots, \b_t)$ in ${\mathbb Z}^t$ let $(a, {\bf \bb})=(a, \b_1,\ldots, \b_t)$ and ${\bf \bb}+a:=(\b_1+a,\ldots, \b_t+a)$.

The next lemma relates the set $\widehat{H}_{\bf \bb}$ with the Weierstrass semigroup $H(P)$ at the place $P$, the generalized Weierstrass semigroups $\widehat{H}({\bf Q}_t)$ and $ \widehat{H}(P,{\bf Q}_t):=\widehat{H}(P, Q_1, \ldots, Q_t)$. 

\begin{lemma}\label{hinhb}
Let ${\bf \bb}=(\b_1,\ldots, \b_t) \in \mathbb{Z}^t$. Then we have
\begin{enumerate}[(i)]
\item $\{a\in \Z :\, (a, {\bf \bb})\in \widehat{H}(P,{\bf Q}_t) \}\subseteq \widehat{H}_{\bf \bb}\subseteq \{a\in\Z  \, :\,  a +(\b_1+\cdots +\b_t) \geq 0\}$ and, if ${\bf \bb}\in \widehat{H}({\bf Q}_t)$ then
$\widehat{H}_{\bf \bb}=\{a\in \Z\, :\, (a, {\bf \bb})\in \widehat{H}(P,{\bf Q}_t) \}.$
\item
 If  ${\bf \bb}\in \widehat{H}({\bf Q}_t)$, then $H(P)\subseteq \{a\in \Z\, :\, (a, {\bf \bb})\in \widehat{H}(P,{\bf Q}_t) \}$.
\item \label{largea} If $a \in \Z$ and $a \geq 2g-(\b_1+\cdots+\b_t)$, then  $a\in \widehat{H}_{\bf \bb}$.
\end{enumerate}
\end{lemma}
\begin{proof}
\begin{enumerate}[(i)]
\item The first inclusion follows by Proposition \ref{Propositiondimension}. For the second inclusion, if $\ell(aP+{\bf G_{\bb}})\neq \ell((a-1)P+{\bf G_{\bb}})$ then $\ell(aP+{\bf G_{\bb}}) \geq 1$ and we obtain $a +(\b_1+\cdots +\b_t) \geq 0$.
Suppose  ${\bf \bb}\in \widehat{H}({\bf Q}_t)$ and let $a\in  \widehat{H}_{\bf \bb}$. Then there exists a function $f\in
 \cL(aP+{\bf G_\bb})\setminus\cL((a-1)P+{\bf G_\bb})$. The pole divisor of
        $f$ is $(f)_\infty=aP+r_1Q_1+\cdots+r_tQ_t$ with $r_i\leq \b_i$ for $i=1,\ldots,t$. This implies that
        $(a, r_1,\ldots,r_t)\in \widehat{H}(P,{\bf Q}_t)$ and as $(0, \b_1,\ldots, \b_t)\in \widehat{H}(P, {\bf Q}_t)$ then from  (\ref{e:lubclosed}) we have  $(a, \b_1,\ldots, \b_t)\in \widehat{H}(P,{\bf Q}_t)$.  
\item Since $H(P)\times \widehat{H}({\bf Q}_t) \subseteq \widehat{H}(P, {\bf Q}_t)$, we get 
  $H(P)\subseteq\{a\in\Z\, :\, (a,{\bf \b})\in \widehat{H}(P,{\bf Q}_t) \}.$
\item  From $a+\b_1+\cdots+\b_t\geq 2g$ we have that $(a,{\bf \bb})\in \widehat{H}(P,{\bf Q}_t)$ and we conclude $\ell(aP+{\bf G_\bb})= \ell((a-1)P+{\bf G_\bb})+1$.
\end{enumerate}
\end{proof}
   
Now, let $D=P_1+ \cdots +P_n$ be the sum of $n$ pairwise distinct
rational places different than $P$ and different than $Q_1,\ldots, Q_t$. 
For ${\bf \bb}=(\b_1,\ldots, \b_t)$ in ${\mathbb Z}^t$ define 
$$
\widehat{H}_{\bf \bb}^*=\{a\in\Z \, :\, C_\mathcal L(D, aP+{\bf G_\bb})\neq C_\mathcal L(D, (a-1)P+{\bf G_\bb}) \}.
$$
We notice that, if $a$ in $\widehat{H}_{\bf \bb}^*$, then $a+(\b_1+\dots +\b_t)\geq 0$. It holds that 
$a \in \widehat{H}_{\bf \bb}^*$ if and only if
$$\dim C_\mathcal L(D, aP+{\bf G_\bb})-\dim C_\mathcal L(D, (a-1)P+{\bf G_\bb})=1.$$

Furthermore, it holds that 
$\# \widehat{H}_{\bf \bb}^*=n$. This is a consequence of the fact that
for 
$a<-\sum_{i=1}^t\beta_i$ one has $\dim(C_\mathcal L(D, aP+{\bf G_\bb}))=0$, while for
$ a = n+2g-1-\sum_{i=1}^t\beta_i$ one has 
$\dim(C_\mathcal L(D, aP+{\bf G_\bb}))=n.$

The next lemma relates the sets $\widehat{H}_{\bf \bb}$ and $\widehat{H}_{\bf \bb}^*$.
\begin{lemma}\label{characterizationhbstar}
For $ {\bf \bb}=(\b_1,\ldots, \b_t) \in {\mathbb Z}^t$ we have
\begin{enumerate}[(i)]
\item $\widehat{H}_{\bf \bb}^* \subseteq \widehat{H}_{\bf \bb}$, and 
\item $\widehat{H}_{\bf \bb} \setminus \widehat{H}_{\bf \bb}^*=\{ a\in\Z \, :\, \ell(aP+{\bf G_\bb}-D) \neq \ell((a-1)P+{\bf G_\bb}-D)\}.$
\item \label{equivHH*}
If $ a + \b_1+\cdots+\b_t < n$,   then $a \in \widehat{H}_{\bf \bb}^*$ if and only if $a \in \widehat{H}_{\bf \bb}$. 
\end{enumerate}
\end{lemma}
\begin{proof}
Item (i) is obvious. Item (ii) follows from equality \eqref{dimcodedivisordifference}. Let us prove now item (iii). We already have $\widehat{H}_{\bf \bb}^* \subseteq \widehat{H}_{\bf \bb}$. Suppose $a \in \widehat{H}_{\bf \bb}$. From $a+\b_1+\cdots+\b_t < n$ and applying equality \eqref{dimcodedivisordifference}, since $\deg((aP+{\bf G_\bb)-D)<0}$,  yields $\dim C_\mathcal L(D, aP+{\bf G_\bb})=\ell(aP+{\bf G_\bb)} \neq \ell((a-1)P+{\bf G_\bb)}=\dim C_\mathcal L(D, (a-1)P+{\bf G_\bb})$.
\end{proof}

From Lemmas \ref{hinhb}(iii)
and \ref{characterizationhbstar}(iii) we can conclude the inclusion $$\left\{2g-\sum_{i=1}^t \b_i, 2g-\sum_{i=1}^t \b_i+1, \dots, n-1-\sum_{i=1}^t \b_i\right\} \subset \widehat{H}_{\bf \bb}^*,$$
which will be used later in the examples.

The next lemma bounds the maximum element in $\widehat{H}_{\bf \b}^*$. For the case of Kummer extensions and algebraic geometry codes, in \cite{BCQ2021}, the authors computed the exact value of the maximum of $H_{\bf b}^*$.

\begin{lemma}
  \label{hbstarbound}
Suppose that the genus of ${\cF}$ is nonzero.
For a general ${\bf \bb}=(\b_1,\ldots, \b_t)\in{\mathbb Z}^t$,
$$
n-(\b_1+\cdots+\b_t) \leq \max(\widehat{H}_{\bf \bb}^*)\leq n+2g-1-(\b_1+\cdots+
\b_t).$$
\end{lemma}
\begin{proof}
  For the first inequality let $a_0\in\Z$ be minimum such that $\dim \cC_\cL(D, aP+{\bf G_\bb})=n,
  $ for all $a \geq a_0$. Since $\dim \cC_\cL(D, a_0P+{\bf G_\bb})
  =\ell(a_0P+{\bf G_\bb})-\ell(a_0P+{\bf G_\bb}-D)$, we have that if $a_0+ \b_1+\cdots+\b_t< n-1$ then
  $\dim \cC_\cL(D,a_0P+{\bf G_\bb}) =\ell(a_0P+{\bf G_\bb}) \leq a_0+\b_1\cdots+\b_t +1<n$, a contradiction. We conclude that $a_0+\b_1+\cdots+\b_t \geq n-1$. If $\max(\widehat{H}_{\bf \bb}^*) = n-(\b_1+\cdots+\b_t)-1$ then $n=\dim \cC_\cL(D, (n-(\b_1+\cdots+\b_t)-1)P+{\bf G_\bb})$ and as $\deg((n-(\b_1+\cdots+\b_t)-1)P+{\bf G_\bb})=n-1<n$ then by equality \eqref{dimcode} we have that $\ell((n-(\b_1+\cdots+\b_t)-1)P+{\bf G_\bb})=n-g$, and we conclude $g=0$.

For the second inequality,  by the Riemman-Roch Theorem, we have 
  $\ell((n+2g-1-(\b_1+\cdots+\b_t))P+{\bf G_\bb}-D)=g$,
  while $\ell((n+2g-1-(\b_1+\cdots+\b_t))P+{\bf G_\bb})=g+n$ because
  $\deg((n+2g-1-(\b_1+\cdots+\b_t))P+{\bf G_\bb}-D)=2g-1+n\geq \deg((n+2g-1-(\b_1+\cdots+\b_t))P+{\bf G_\bb}-D)\geq  2g-1$. On the other hand, by equality
\eqref{dimcodedivisordifference}, 
  \begin{align*}
 & \dim C_\mathcal{L}(D, (n+2g-1-\sum\limits_{i=1}^t\b_i)P+{\bf G_\bb})=\\
  &\ell((n+2g-1-\sum\limits_{i=1}^t\b_i)P+{\bf G_\bb})-\ell((n+2g-1-\sum\limits_{i=1}^t\b_i)P+{\bf G_\bb}-D)=n.
\end{align*}
  \end{proof}

\section{The isometry-dual property for many-point codes}

The next definition was first introduced in \cite{GMRT2011}. 
\begin{definition}
A flag of codes $(C_i)_{i=0,\ldots, s}$ is said to satisfy the isometry-dual condition if there exists ${\bf x}\in (\mathbb{F}_q^*)^n$ such that $C_i$ is {\bf x}-isometric to $C_{s-i}^\bot$ for all $i=0,\ldots, s$. 
\end{definition}

\begin{theorem}\label{t:isodual}
Let $\cF$  be a function field of genus $g$ over $\mathbb{F}_q$ and let $P, Q_1,\ldots,Q_t$ be pairwise distinct rational places in $\Pla(\cF)$.
Consider ${\bf \bb}=(\b_1,\ldots, \b_t) \in \Z^t$, and the divisor $D=P_1+ \cdots +P_n$ the sum of $n\geq 2g+2$
distinct
rational places, $P, Q_1,\ldots, Q_t \notin \supp(D)$. 
The following are equivalent.
\begin{enumerate}[(i)]
\item There exists a constant vector $\bf x$ such that the flag of codes
\begin{equation*}\label{seqcodes}
\{0\}=C_\mathcal L(D, a_0P+{\bf G_\bb})\subsetneq
C_\mathcal L(D, a_1P+{\bf G_\bb})\subsetneq
C_\mathcal L(D, a_2P+{\bf G_\bb})\subsetneq
\dots \subsetneq C_\mathcal L(D, a_nP+{\bf G_\bb})=\mathbb{F}_{q}^n
\end{equation*} 
is $\bf x$-isometry-dual,
where $-1-\sum_{i=1}^t \beta_i\leq a_0=a_1-1$
and $a_1<\dots < a_{n}$ is the ordered set of elements in $\widehat{H}_{\bf \bb}^*$.
\item
The divisor $E=(n+2g-2-2(\b_1+\cdots+\b_t))P+2{\bf G_\bb}-D$ is canonical.
\item
$n+2g-1-2(\b_1+\cdots +\b_t) \in \widehat{H}_{2{\bf \bb}}^*$.
\end{enumerate}
\end{theorem}
\begin{proof}
Let us first prove that (i) implies (ii). Since $n\geq 2g+2$, we can take $a$  an integer such that
$$2g-\b_1-\cdots-\b_t\leq a \leq n-2-\b_1-\cdots-\b_t.$$
By Lemma~\ref{hinhb}(iii), we get $a\in  \widehat{H}_{\bf \bb}$. Furthermore, since $a \leq n-2-(\b_1+\cdots+\b_t)$,
from Lemma~\ref{characterizationhbstar}(iii) we conclude
$a\in \widehat{H}_{\bf \bb}^*$.
Define $a^\perp=n+2g-2-2(\b_1+\cdots+\b_t)-a$. As before, since $a\leq n-2-(\b_1+\cdots+\b_t)$, we have
$a^\perp\geq 2g-(\b_1+\cdots+\b_t)$ and hence $a^\perp\in \widehat{H}_{\bf \bb}$. Furthermore, since $a\geq
2g-(\b_1+\cdots+\b_t)$ we also have $a^\perp\leq n-2-(\b_1+\cdots+\b_t)$, and so $a^\perp \in \widehat{H}_{\bf \bb}^*$.

Notice that, since $2g\leq a+\b_1+\cdots+\b_t,a^\perp+\b_1+\cdots+\b_t<n$, and by equality \eqref{dimcode},
it holds $\dim(C_{\mathcal 
  L}(D, aP+{\bf G_\bb}))=a+\b_1+\cdots+\b_t+1-g$ and $\dim(C_{\mathcal 
  L}(D, a^\perp P+{\bf G_\bb}))=a^\perp+\b_1+\cdots+\b_t+1-g=n-\dim(C_{\mathcal 
  L}(D, aP+{\bf G_\bb}))$.
So, the codes associated to $a$ and $a^\perp$ correspond to dual codes in the isometry-dual flag.  
 
From equation \eqref{dualcode} we have that $C_\cL(D, aP+{\bf G_\bb})^\perp$ is $C_\cL(D, D+W-aP-{\bf G_\bb})$,
where $W$ is a canonical divisor  
with $v_{P_i}(W)=-1$ for any $P_i$ in $\supp(D)$.
Assuming the ${\bf x}$-isometry-dual property we deduce that
$D+W-aP-{\bf G_\bb}\sim a^\perp P+{\bf G_\bb}$, since $a \geq 2g -(\b_1+\cdots+\b_t)$.
Then $W\sim (a+a^\perp)P+2{\bf G_\bb}-D=(n+2g-2-2(\b_1+\cdots+\b_t))P+2{\bf G_\bb}-D=E$.
Hence $E$ is canonical.

Now we prove that (ii) implies (i).
Suppose that $E=(n+2g-2-2(\b_1+\cdots+\b_t))P+2{\bf G_\bb}-D$ is a canonical
divisor. Let $W$ be a canonical divisor  with $v_{P_i}(W)=-1$ for any $P_i$ in $\supp(D)$ (see equation \eqref{dualcode}).
Then there is a rational function $f$ such that $E+(f)=W$. 
In particular, $f$ has neither poles, nor zeros in the support of $D$.
Let ${\bf x}=ev_D(f)$. Then, for any $a\in \widehat{H}_{\bf \bb}^*$,
let $a^\perp=n+2g-2-2(\b_1+\cdots+\b_t)-a$
and $(a^\perp)^*=\max\{a \in \widehat{H}_{\bf \bb}^*: a\leq a^\perp\}$.
Notice that $(a^\perp)^*\in \widehat{H}_{\bf \bb}^*$ and
$C_\cL(D, (a^\perp)^* P+{\bf G_\bb})= C_\cL(D, a^\perp P+{\bf G_\bb})$.
We have 
\begin{align*}
D+W-(aP+{\bf G_\bb})&=D+E+(f)-(aP+{\bf G_\bb})\\
 & =
(n+2g-2-2(\sum\limits_{i=1}^t\b_i))P+2{\bf G_\bb}+(f)-(aP+{\bf G_\bb})\\
& =
(a^\perp P+{\bf G_\bb})+(f).
\end{align*}
Hence, $C_\cL(D, aP+{\bf G_\bb})^\perp={\bf x}\cdot C_\cL(D, a^\perp P+{\bf G_\bb})={\bf x}\cdot C_\cL(D, (a^\perp)^* P+{\bf G_\bb})$.
With this we proved that, assuming (ii), there exists a vector ${\bf x}$ such that the dual code of any code of the form $C_\cL(D, aP+{\bf G_\bb})$, where  $a\in \widehat{H}_{\bf \bb}^*$
is exactly ${\bf x}\cdot C_\cL(D, a' P+{\bf G_\bb})$
for some $a'\in \widehat{H}_{\bf \bb}^*$.
Hence, the flag in (i) satisfies the isometry-dual property.

Now we prove that (ii) and (iii) are equivalent. 
By the Riemann-Roch Theorem we know that $\ell(E+P)=g$ and, consequently, $\ell(E)\leq g$. Since $\deg(E)=2g-2$, $E$ is canonical if and only if $\ell(E)=g$ (see \cite[Proposition 1.6.2]{St2009}).
Then $E$ is canonical if and only if $\ell(E)=\ell(E+P)$, that is, if and only if 
$$\ell((n+2g-2-2\sum\limits_{i=1}^t\b_i)P+2{\bf G_\bb}-D)=\ell((n+2g-1-2\sum\limits_{i=1}^t\b_i)P+2{\bf G_\bb}-D).$$ By Lemma~\ref{characterizationhbstar}(ii) this is equivalent to $n+2g-1-2(\b_1+\cdots+\b_t)\in \widehat{H}^*_{2{\bf \bb}}$.
\end{proof}

  \begin{example}\label{klein}
    Let ${\mathbb F}_8={\mathbb F}_2(\alpha)$ with $\alpha^3+\alpha+1=0$.
    Consider the Klein function field over ${\mathbb F}_8$
    defined by the curve ${\mathcal X}$ with equation $X^3Y+Y^3Z+XZ^3=0$.
This curve has 24 rational places and genus $g= 3 $. Take 
$P=( 0 : 0 : 1 )$,
and $Q_1=(0:1:0), Q_2=(1:0:0)$ the two places at infinity, and let 
$D$ the sum of the $n=21$ rational places of ${\mathcal X}$ except $P$, $Q_1$, and $Q_2$.
Given the divisor $G=aP+\beta_1 Q_1+3Q_2$, for each $\beta_1$ we consider the flag
$$S_{(\beta_1,3)}:=\left(C_{\mathcal L}(D,aP+\beta_1 Q_1+3Q_2)\right)_{a\in \widehat H_{(\beta_1,3)}^*\cup\{-1+\min \widehat H_{(\beta_1,3)}^*\}}.$$
The values of $\beta_1$ for which $S_{(\beta_1,3)}$ satisfies the isometry-dual property are exactly the values in the set $\{-18,-11,-4,3,10,17,\dots\}$.
The corresponding flags of codes $S_{(-18,3)}$, $S_{(-11,3)}$, $S_{(-4,3)}$, $S_{(3,3)}$, $S_{(10,3)}$, $S_{(17,3)}$, are remarked with circles in
Figure~\ref{fig:QQDKleinisometryduallarge},
where a circle at position $(a,\beta_1)$ refers to the code $C_{\mathcal L}(D,aP+\beta_1 Q_1+3Q_2)$ and flags correspond to the circles in a horizontal line. We computed the vector $x$ giving the isometry-dual property for each flag satisfying it.
In Figure~\ref{fig:QQGKleinisometryduallarge}
one can see the sets $\widehat H^*_{(\tilde \beta_1,6)}$ and its upper bound
$n+2g-1-(\tilde \beta_1+6)$.
The horizontal lines of stars intersecting the diagonal line, correspond to the values $(\tilde \beta_1,6)$ for which
$n+2g-1-(\tilde \beta_1+6) \in \widehat{H}_{(\tilde \beta_1,6)}^*$.
If intersection occurs and $\tilde \beta_1=2\beta_1$ for some $\beta_1$ (i.e. if $\tilde \beta_1$ is even), then this means that $S_{(\beta_1,3)}=S_{(\tilde \beta_1/2,3)}$ satisfies the isometry dual property, and this can be ckecked looking again at Figure~\ref{fig:QQDKleinisometryduallarge}.
\end{example}

\newcommand\graphkleinD{}
\newcommand\graphkleinG{}

\newcommand\graphkleinA{}
\newcommand\graphkleinB{}
\newcommand\graphkleinC{}
\newcommand\graphkleinE{}
\newcommand\graphkleinF{}
\newcommand\graphkleinH{}
\newcommand\graphkleinI{}
\newcommand\graphkleinJ{}
\newcommand\graphkleinK{}
\newcommand\graphkleinV{}



\newcommand\marca{{\thicklines\circle{1.01}\circle{1.}\circle{.99}}}
\newcommand\sensemarca{}

\renewcommand\graphkleinD{
  \begin{figure}[h]
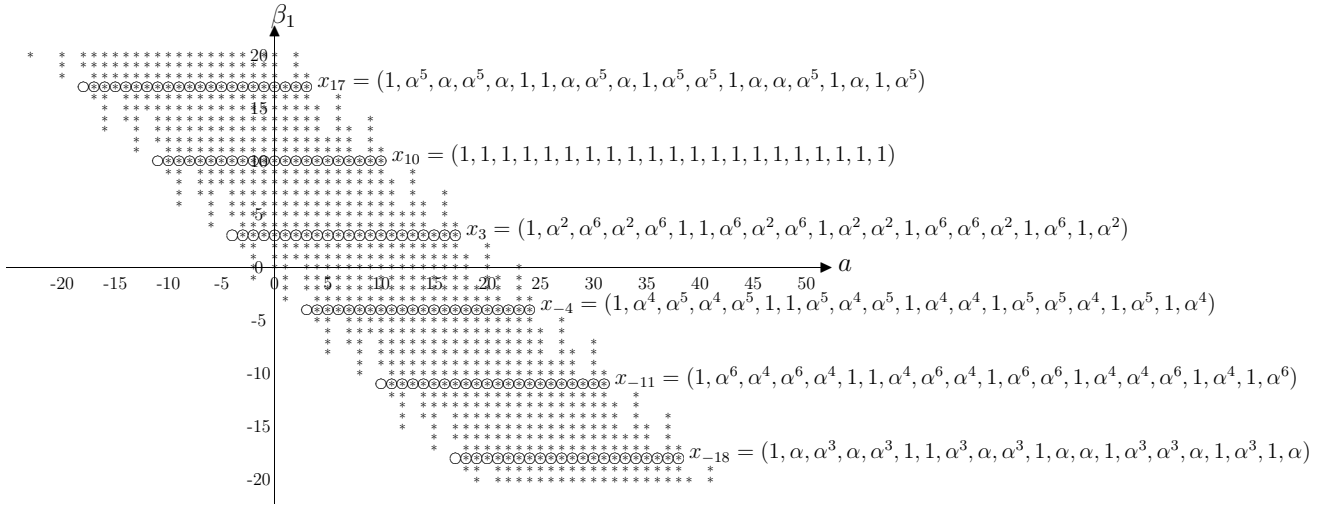

  \caption{ Klein  function field over ${\mathbb F}_{ 8 }$.
Analysis of the codes $C_{{\mathcal L}}(D,aP+\beta_1 Q_1+3Q_2)$.
}
\setlength{\unitlength}{1cm}
\label{fig:QQDKleinisometryduallarge}
\begin{minipage}{ 0.650000000000000 \textwidth}
\resizebox{\textwidth}{!}{

at $(a,\beta_1)$ if $C_{{\mathcal L}}(D,aP+\beta_1 Q_1+3Q_2)$
in an isometry-dual flag $\Longleftrightarrow$\\
\phantom{M}$n+2g-1-2(\beta_1+3)=20-2\beta_1\in \widehat H^*_{(2\beta_1,6)}$ in Fig.~\ref{fig:QQGKleinisometryduallarge}
(by Theorem~5.5.)
}\end{minipage}\end{minipage}
  \end{figure}
}
\renewcommand\graphkleinG{
  \begin{figure}
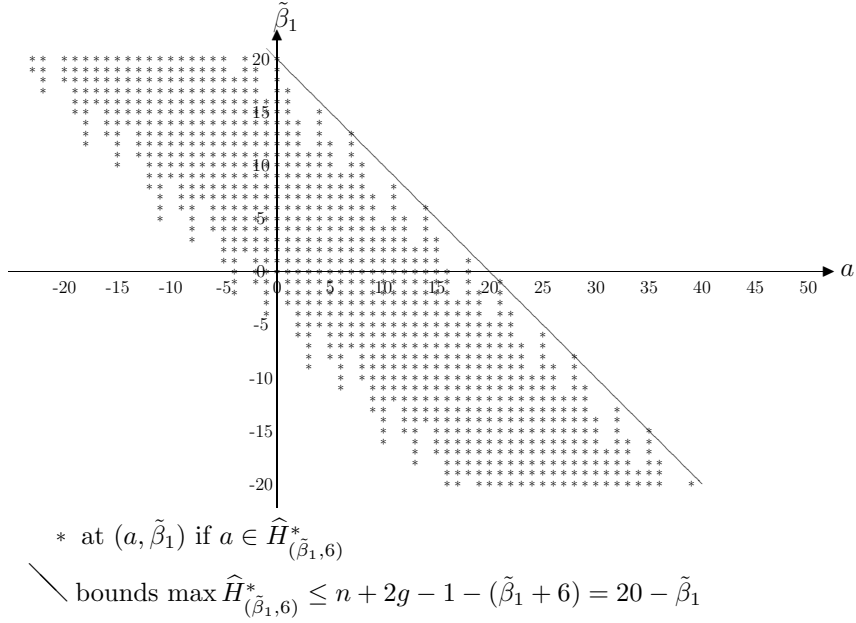

  \caption{ Klein  function field over ${\mathbb F}_{ 8 }$.
    Analysis of
    the sets $\widehat H^*_{(\tilde \beta_1,6)}$.
}

\setlength{\unitlength}{1cm}
\label{fig:QQGKleinisometryduallarge}
\begin{center}
\begin{minipage}{ 0.650000000000000 \textwidth}
\resizebox{\textwidth}{!}{

bounds $\max \widehat H_{(\tilde\beta_1,6)}^* \leq n+2g-1-(\tilde\beta_1+6)=20-\tilde\beta_1$
}\end{minipage}\end{minipage}
\end{center}\end{figure}
}

\graphkleinD
\graphkleinG

\begin{corollary}\label{computationHbetastar}
  With the same notation as in Theorem \ref{t:isodual}, if the flag
\begin{equation*}
\{0\}=C_\mathcal L(D, a_0P+{\bf G_\bb})\subsetneq
C_\mathcal L(D, a_1P+{\bf G_\bb})\subsetneq
C_\mathcal L(D, a_2P+{\bf G_\bb})\subsetneq
\dots \subsetneq C_\mathcal L(D, a_nP+{\bf G_\bb})=\mathbb{F}_{q}^n
\end{equation*} 
  satisfies the isometry-dual condition, then $$\widehat{H}_\bb^*  =\{a\in \widehat{H}_\bb: a^\bot+1\in \widehat{H}_\bb\},$$
where $a^\bot=n+2g-2-2(\b_1+\cdots+\b_t)-a$.
\end{corollary}
\begin{proof}
  Let $a\in \widehat{H}_\bb$. From the Riemann-Roch Theorem, we have that 
$\ell(aP+{\bf G_\bb}-D)=a+\b_1+\cdots+\b_t-n+1-g+\ell(W+D-aP-{\bf G_\bb}),$
  where $W$ is a canonical divisor. From the isometry dual condition, applying Theorem \ref{t:isodual},
since the divisor $E=(n+2g-2-2\sum_{i=1}^t\b_i)P+2{\bf G_\bb}-D$ is canonical, we have the following equivalence of divisors $D+W-aP-{\bf G_\bb}\sim a^\bot P+{\bf G_\bb}$.
Now, by Lemma~\ref{characterizationhbstar}(ii),
\begin{align*}
  &a\in \widehat{H}_\bb^*
\Leftrightarrow \ell(aP+{\bf G_\bb}-D)=\ell((a-1)P+{\bf G_\bb}-D)\\
  &\Leftrightarrow {\small a+\sum_{i=1}^t\b_i-n+1-g+\ell(E+D-aP-{\bf G_\bb})= a-1+\sum_{i=1}^t \b_i-n+1-g+\ell(E+D-(a-1)P-{\bf G_\bb})}\\
  &\Leftrightarrow   \ell(E+D-(a-1)P-{\bf G_\bb})-\ell(E+D-aP-{\bf G_\bb})= 1\\
  &\Leftrightarrow \ell((n+2g-2-2\sum_{i=1}^t\b_i)P+{\bf G_\bb}-(a-1)P)-\ell((n+2g-2-2\sum_{i=1}^t\b_i)P+{\bf G_\bb}-aP)= 1\\
  &\Leftrightarrow \ell((n+2g-2-2\sum_{i=1}^t\b_i-a)+1)P+{\bf G_\bb})-\ell((n+2g-2-2\sum_{i=1}^t\b_i-a)P+{\bf G_\bb})= 1\\
  &\Leftrightarrow \ell(a^\perp+1)P+{\bf G_\bb})-\ell(a^\perp P+{\bf G_\bb})= 1
\Leftrightarrow a^\bot+1\in \widehat{H}_\bb.
\end{align*}
\end{proof}

\section{The isometry-dual property for many-point codes over Kummer extensions}
\label{s:K}

 Let $\cF/\fq$ be a  Kummer extension defined by the affine equation 
 \begin{equation}\label{Kummer}
 y^m=f(x)=\prod\limits_{i=1}^{r}(x-\alpha_i), \quad \alpha_i \in \fq
 \end{equation}
 where $f(x)$ is a separable polynomial of degree $2\leq r\leq m-1$ with $\gcd(m, r)=1$. This function field has genus $g=(m-1)(r-1)/2$ and only one rational place at infinity $P_\infty$, the common pole of
$x$ and $y$. Let $R_\infty \in \Pla(\fq(x))$ be the only pole of $x$, and $R_1, \dots, R_r$ stand for the places of the rational function field $\fq(x)$ associated to the zeros $\alpha_1,\ldots, \alpha_r$ of $f(x)$,  respectively. Since the places $R_1, \dots, R_r$ are totally ramified in the extension $\cF/\fq(x)$, there exists a unique rational place $P_{\alpha_i,0}$ in $\Pla(\cF)$ over $R_i$ for $i=1, \dots ,r$. Let $R_{r+1}, \dots ,R_{r_1}$ be all the rational places in $\Pla(\fq(x))$ completely split in the extension $\cF/\fq(x)$. Notice that $P_\infty$ in $\Pla(\cF)$ is the only place over $R_\infty$. Then such a function field $\cF$ has exactly $r+1+(r_1-r)m$ rational places in total.

From now on in this section,
fix  $Q_1=P_{\alpha_1,0}, Q_2=P_{\alpha_2,0}, \dots , Q_t=P_{\alpha_t,0}, 1\leq t \leq r$.
In $\cF$ we  have the following divisors :
\begin{enumerate}[(i)]
\item $(x-\alpha_i)=mQ_i-mP$  for every $i$, $1\le i \le r$,
\item $(y)= Q_1+\cdots +  Q_r-r P$.
\end{enumerate}
Let $D$ be the sum of $n$ rational places in $\Pla(\cF)$ such that $P, Q_i \not\in \supp(D), i=1, \dots t$. We investigate the dimension of the many-point codes $C_\cL(D, aP+{\bf G_ \bb})$ where ${\bf G_ \bb}=\b_1Q_1+\cdots+\b_tQ_t$ for $a, \b_i \in \Z$.

For any $J\subseteq I=\{1,\ldots,t\}$, we set $J^c=I\setminus J$ and denote by ${\bf 1}_J$ the $t$-tuple having the $j$-th coordinate 1 if $j\in J$, and $0$ otherwise. Let $e_i={\bf 1}_{\{i\}}$.  Now we notice that given $\b_i \in \Z$, we can write  $\b_i=\theta_i m+\b_i'$ with $ 0\leq \b_i'<m, \theta_i \in \Z$.

\begin{remark}\label{r:smallbetas}
Multiplication by the function $(x-\alpha_i)^{\theta_i}$ gives an isomorphism between the  Riemann-Roch spaces $\cL( aP+{\bf G}_\bb)$ and  $\cL((a+\theta_i m)P+{\bf G}_{\bb-\theta_i me_i})$, and so the codes $C_\cL(D, aP+{\bf G}_\bb)$ and $C_\cL(D, (a+\theta_i m)P+{\bf G}_{\bb-\theta_i me_i})$ are isometric. 
For $i=1, \dots , t$ writing $\b_i=\theta_im+\b_i'$ with $ 0\leq \b_i'<m$ yields
$$a \in \widehat{H}_{\bf \bb}^* \mbox{ if and only if } a+(\theta_1+\cdots+\theta_t)m \in \widehat{H}_{\bf \bb'}^*$$ 
where $\bb'=(\b_1',\ldots,\b_t')$. 
\end{remark}

Now we determine necessary and sufficient conditions for an integer $a$ to be in $ \in \widehat{H}_{\bf \bb}^*$. At first we deal with the case in which the place $P_\infty$ is in the support of the divisor $G$.

\begin{theorem}\label{t:kummer}
Let $\mathcal{F}/\fq$ be a Kummer extension of genus $g$  given by 
$$y^m=f(x)=\prod\limits_{i=1}^{r}(x-\alpha_i), $$  
where $\alpha_i \in \fq ,$  $f(x)$ is a separable polynomial of degree $2\leq r \leq m-1$, and $\gcd(m, r)=1$. 
Let $R_{r+1}, \dots ,R_{r_1}$ be rational places in $\Pla(\fq(x))$ completely split in the extension $\cF/\fq(x)$.
Let $P=P_\infty $ and $Q_1=P_{\alpha_1,0}, Q_2=P_{\alpha_2,0}, \dots , Q_t=P_{\alpha_t, 0}, 1\leq t \leq r$.
Consider the divisors
$$G= aP+{\b_1Q_1+\cdots+\b_tQ_t}=aP+{\bf G_\bb},\ \ \ \ a, \b_i \in \Z, $$
and 
$$D=\sum\limits_{i=t+1}^{r} Q_i + \sum\limits_{i=r+1}^{r_1} \sum\limits_{\tilde P\mid R_i} \tilde P \quad  \text{an } Aut(\cF/\fq(x))\text{-invariant divisor.}$$
Denote $n=\deg(D)$ and suppose $n\geq 2g.$ Let $\tilde r$ be the inverse of $r$ mod $m$.
Then, $a\in \widehat{H}_{\bf \bb}^*$ if and only if
\begin{enumerate}[i)]
\item $\sum \limits_{i=1}^{t} \left \lfloor \frac{\b_i+j}{m}  \right\rfloor\geq \frac{rj-a}{m}$, where $j$ is the residue of $\tilde r a$ modulo $m$,
        if  $0 \leq a+\sum_{i=1}^t \b_i<n.$
        
\item $\sum \limits_{i=1}^{t} \left \lfloor \frac{\b_i+j+1}{m}  \right\rfloor<\frac{rj+n+t-a}{m}$,
    where $j$ is the residue of $\tilde r (a-n-t)$ modulo $m$, 
    if $n\leq a+\sum_{i=1}^t \b_i\leq  n+2g-1.$
\end{enumerate} 
\end{theorem}
\begin{proof}
We use the same notations as above.   
We have $n=r-t+(r_1-r)m\geq (m-1)(r-1)-1$. By equality \eqref{dimcodedivisordifference},
$a \in \widehat{H}_{\bf \bb}^*$ if and only if 
\begin{align}
1&=\ell(aP+{\bf G_\bb})- \ell((a_1-1)P+{\bf G_\bb})
-\ell(aP+{\bf G_\bb}-D)+\ell((a-1)P+{\bf G_\bb}-D).
\end{align}
Hence, $a \in \widehat{H}_{\bf \bb}^*$ if and only if 
\begin{align}\label{dim1}
&\ell(aP+{\bf G_\bb})- \ell((a_1-1)P+{\bf G_\bb})=1
\end{align}
and 
\begin{align}\label{dim0}
&\ell(aP+{\bf G_\bb}-D)-\ell((a-1)P+{\bf G_\bb}-D)=0.
\end{align}
In the same way, denote $R_1, \dots , R_r$ the places in the rational function field $\fq(x)$ associated to the zeros $\alpha_1,\ldots, \alpha_r$ of $f(x)$,
and denote $R_{r+1}, \dots ,R_{r_1}$ the rational places in $\Pla(\fq(x))$ associated to the zeros of the functions $x- \alpha_{r+1}, \ldots, x- \alpha_{r_1} \in \fq(x)$, respectively. Then the divisor of the function $z=y\prod\limits_{i=r+1}^{r_1}(x-\alpha_i)$ in $\cF$ is 
$$(z)=D+Q_1+\dots +Q_t -(n+t)P.$$
This yields the following equivalence of divisors $D\sim  (n+t)P-Q_1-\dots -Q_t$, which allows to conclude 
\begin{equation}\label{equivdivisor}
\ell(aP+{\bf G_\bb}-D)=\ell((a-(n+t))P+{\bf G_{\bb+1}}).
\end{equation}	
By \cite[Theorem 2.2]{M2004}, we have that 
\begin{align*}
\cL(aP+{\bf G_\bb})=\bigoplus\limits_{j=0}^{m-1}  \mathcal{L}([aP+{\bf G_\bb}+(y^j)]_{|\fq(x)})\,y^j,
\end{align*} 
which gives 
\begin{eqnarray*}
&\ell(aP+{\bf G_\bb})&=\sum \limits_{j=0}^{m-1}  \ell([aP+{\bf G_\bb}+(y^j)]_{|\fq(x)})\\
&&=\sum \limits_{j=0}^{m-1}  \ell([aP+{\bf G_\bb}+\sum\limits_{i=1}^{r} j Q_i-rj P]_{|\fq(x)})\\
&&=\sum \limits_{j=0}^{m-1}  \ell\left(\left\lfloor \frac{a-rj}{m} \right\rfloor R_\infty +\sum \limits_{s=1}^{t} \left \lfloor \frac{\b_s+j}{m}  \right\rfloor R_s \right).
\end{eqnarray*} 
For $j=0, \ldots , m-1$ and ${\bf \bb}=(\b_1, \dots \b_t)$ define $D_{j, a, {\bf \bb}}=\left\lfloor \frac{a-rj}{m} \right\rfloor R_\infty +\sum \limits_{s=1}^{t} \left \lfloor \frac{\b_s+j}{m}  \right\rfloor R_s .$
Then, using equation \ref{equivdivisor}  we can conclude that $a \in \widehat{H}_{\bf \bb}^*$ if and only if 
\begin{align*}
&\sum \limits_{j=0}^{m-1} \left[\ell( D_{j, a, {\bf \bb}})- \ell( D_{j, a-1, {\bf \bb}})\right] =\ell(aP+{\bf G_\bb})- \ell((a-1)P+{\bf G_\bb})=1 \text{ and }\\
&\sum \limits_{j=0}^{m-1} \left[\ell (D_{j, a-(n+t), {\bf \bb+1}})- \ell (D_{j, a-(1+n+t), {\bf \bb+1}})\right]=\ell(aP+{\bf G_\bb}-D)-\ell((a-1)P+{\bf G_\bb}-D)=0.
\end{align*} 

In the rational function field $\fq(x)$ for any divisor $A$ with $\deg(A) \geq -1$ we have $\ell(A)=\deg(A)+1$. This yields two cases to be analyzed depending on the value of the sum $\deg(aP+{\bf G_\bb})=a+\b_1+\cdots +\b_t$. 

{\bf Case 1:} if $0\leq \deg(aP+{\bf G_\bb}) <n$ then $\ell(aP+{\bf G_\bb}-D)=\ell((a-1)P+{\bf G_\bb}-D)=0$, and we are left to analyze whether the difference of dimensions $\ell(aP+{\bf G_\bb})- \ell((a-1)P+{\bf G_\bb})$ is equal to one. We have
$\sum \limits_{j=0}^{m-1} \left[\ell(D_{j, a, {\bf \bb}})- \ell(D_{j, a-1, {\bf \bb}})\right]=1$ if and only if the following two conditions are satisfied
\begin{enumerate}[i)]
\item $\exists\,\, 0 \leq j_0 \leq m-1 \text{ such that } \deg(D_{j_0, a-1, \bb})  \geq -1  \text{, and  }$
\item $\left\lfloor \frac{a-rj_0}{m} \right\rfloor=\left\lfloor \frac{a-1-rj_0}{m} \right\rfloor+1.$
\end{enumerate}

{\bf Case 2: } if $\deg(aP+{\bf G_\bb})\geq n>2g-1$, then by the Riemann-Roch theorem
we have $\ell(aP+{\bf G_\bb})- \ell((a-1)P+{\bf G_\bb})=1$, and we are left to analyze the condition $\ell(aP+{\bf G_\bb}-D)-\ell((a-1)P+{\bf G_\bb}-D)=0$. In this case,
$\sum \limits_{j=0}^{m-1} \left[\ell (D_{j, a-(n+t), {\bf \bb+1}})- \ell (D_{j, a-(1+n+t), {\bf \bb+1}})\right]=0$ if and only if for every $j  \in \{0, \dots, m-1\}$ one of the following conditions is satisfied:
\begin{enumerate}[i)]
\item $\deg(D_{j, a-(n+t), {\bf \bb+1}})=\left\lfloor \frac{a-rj-(n+t)}{m} \right\rfloor +\sum \limits_{s=1}^{t} \left \lfloor \frac{\b_s+j+1}{m}  \right\rfloor  <0$ or \\
\item $\deg(D_{j, a-(n+t), {\bf \bb+1}})  \geq 0$  and $ \left\lfloor \frac{a-rj-(1+n+t)}{m} \right\rfloor=\left\lfloor \frac{a-rj-(n+t)}{m} \right\rfloor.$
\end{enumerate} 
These conditions can be resumed as  $a \in \widehat{H}_\bb^*$  if and only if 
\begin{enumerate}[i)]
\item $\exists\,\, 0 \leq j_0 \leq m-1 \text{ such that } m|(a-rj_0)$ and $ \frac{a-rj_0}{m} +\sum \limits_{s=1}^{t} \left \lfloor \frac{\b_s+j_0}{m}  \right\rfloor \geq 0$, if  $0 \leq a+\b_1+\cdots + \b_t<n.$ 
\item for every $0 \leq j \leq m-1, \left\lfloor \frac{a-rj-(n+t)}{m} \right\rfloor +\sum \limits_{s=1}^{t} \left \lfloor \frac{\b_s+j+1}{m}  \right\rfloor  <0$ or $m \nmid  a-rj-(n+t)$, if $n\leq a+\b_1+\cdots +\b_t\leq  n+2g-1.$
\end{enumerate} 
\end{proof}

By the definition $\widehat H_\b$ and from the proof of Theorem \ref{t:kummer} we also have the following corollary.
\begin{corollary}\label{ainHbeta} In particular, with the same notation and hypotheses as in Theorem \ref{t:kummer} we have that $a \in \widehat H_\b$ if and only if
\begin{enumerate}[i)]
\item $ a \geq n-\sum_{i=1}^t \b_i$ or 
\item $-\sum_{i=1}^t \b_i \leq a < n-\sum_{i=1}^t \b_i$ and 
  $ \frac{rj_0-1}{m} \leq \sum \limits_{i=1}^{t} \left \lfloor \frac{\b_i+j_0}{m}  \right\rfloor $,
  where $j_0$ is the residue of $a\tilde r$ modulo $m$.
  \end{enumerate}
\end{corollary}

We now deal with the case in which the place $P_\infty$ is not in the support of the divisor $G$.

\begin{theorem}\label{t:kummerseminfinito}
Let $\mathcal{F}/\fq$ be a Kummer extension of genus $g$  given by 
$$y^m=f(x)=\prod\limits_{i=1}^{r}(x-\alpha_i), $$  
where $\alpha_i \in \fq ,$  $f(x)$ is a separable polynomial of degree $2\leq r \leq m-1$, and $\gcd(m, r)=1$. 
Let $R_{r+1}, \dots ,R_{r_1}$ be rational places in $\Pla(\fq(x))$ completely split in the extension $\cF/\fq(x)$.
Let $P=Q_1 $ and $ Q_2=P_{\alpha_2,0}, \dots , Q_t=P_{\alpha_t, 0}, 1\leq t \leq r$.
Consider the divisors
$$G= aP+{\b_2Q_2+\cdots+\b_tQ_t}=aQ_1+{\bf G_\bb},\ \ \ \ a, \b_i \in \Z$$
and 
$$D=\sum\limits_{i=t+1}^{r} Q_i + \sum\limits_{i=r+1}^{r_1} \sum\limits_{\tilde P\mid R_i} \tilde P  \quad  \text{an } Aut(\cF/\fq(x))\text{-invariant divisor.} .$$
Denote $n=\deg(D)$ and suppose $n\geq 2g.$ Then, $a\in \widehat{H}_{\bf \bb}^*$ if and only if
\begin{enumerate}[i)]
\item  $\sum \limits_{i=2}^{t} \left \lfloor \frac{\b_i+j}{m}  \right\rfloor \geq \left \lfloor \frac{rj}{m}  \right\rfloor -\frac{a+j}{m}$, where $j$ is the residue of $-a$ modulo $m$, if  $0 \leq a+\sum_{i=2}^t \b_i<n.$ 
   
\item  $\sum \limits_{i=2}^{t} \left \lfloor \frac{\b_i+j+1}{m}  \right\rfloor < \left \lfloor \frac{n+t+rj}{m}  \right\rfloor -\frac{a+j+1}{m}, $
    where $j$ is the residue of $-a-1$ modulo $m$, if $n\leq a+\sum_{i=2}^t \b_i\leq  n+2g-1.$ 
\end{enumerate} 
\end{theorem}
\begin{proof} The proof is analogous to the proof of the previous Theorem \ref{t:kummer} and we omit it. 
\end{proof}

We now apply the previous results to decide when a flag of many-point codes defined on a Kummer extension has the isometry dual property. First we analyze the case in which $P_\infty $ is in the support of the divisor $G$.

\begin{theorem}\label{t:isodualKummer2}
With the same notation as in Theorem \ref{t:kummer}, where $P=P_\infty$,
and $ 1\leq t \leq r$, let  $\bb=(\b_1, \dots, \b_t) \in \Z^t$ be such that $\sum_{i=1}^t \b_i< (n+2g-1)/2$.
There exists a constant vector $\bf x$ such that the flag of codes
\begin{equation*}
\{0\}=C_\mathcal L(D, a_0P+{\bf G_\bb})\subsetneq
C_\mathcal L(D, a_1P+{\bf G_\bb})\subsetneq
C_\mathcal L(D, a_2P+{\bf G_\bb})\subsetneq
\dots \subsetneq C_\mathcal L(D, a_nP+{\bf G_\bb})=\mathbb{F}_{q}^n
\end{equation*} 
is $\bf x$-isometry-dual,
where $-1-\sum_{i=1}^t \beta_i\leq a_0=a_1-1$ and $a_1<\dots < a_{n}$ is the ordered set of elements in $\widehat{H}_{\bf \bb}^*$, if and only if
$$
\begin{cases}
m \mid (2\b_i +1) &\text{ for all } i=1,\ldots, t, \text{ if } t<r.\\
2\b_i \equiv 2\beta_j \mod m  &\text{ for all } 1\leq i, j \leq t, \text{ if } t=r.
\end{cases}
$$
In particular, for $m$  even and $t<r$, there is no ${\bf x}$-isometry-dual flag.
\end{theorem}
\begin{proof}
  By Theorem~\ref{t:isodual}, the flag is ${\bf x}$-isometry-dual for some vector ${\bf x}$ if and only if
  $c:=n+2g-1-2\sum_{i=1}^t \b_i\in \widehat{H}_{2\bb}^*$.
 And by Theorem~\ref{t:kummer}(ii),
  $c \in \widehat{H}_{2\b}^*$ if and only if  
\begin{equation}\label{eq:maa}\sum \limits_{i=1}^{t} \left \lfloor \frac{2\b_i+j+1}{m}  \right\rfloor<\frac{rj+n+t-c}{m},\end{equation}
 where $j$ is the residue of $\tilde r (c-n-t)$ modulo $m$. For $i=1, \dots, t$ write $2\b_i+1=mq_i+r_i, 0\leq r_i \leq m-1, q_i \in \Z $. We will use the next equality.
 \begin{eqnarray*}
  c-n-t&
 =& 2g-1-2\sum_{i=1}^t\beta_i-t\\
& =& mr-m-r-m\sum_{i=1}^tq_i-\sum_{i=1}^tr_i\\
 &=&m (r-1-\sum_{i=1}^tq_i)-(r+\sum_{i=1}^tr_i)\\
 \end{eqnarray*}
 On one hand, 
 \begin{equation}
\label{eq:jequiv}
   j\equiv \tilde r(c-n-t)\equiv -\tilde r(r+\sum_{i=1}^tr_i)\equiv -1 -\tilde r \sum_{i=1}^tr_i \mod m
 \end{equation}
 Similarly, we deduce that the right hand side of inequality \eqref{eq:maa} is 
 \begin{eqnarray*}
   \frac{rj+n+t-c}{m}&=& \frac{rj+r+\sum_{i=1}^tr_i}{m}+\sum_{i=1}^t q_i -r+1\\
 \end{eqnarray*}
 The left hand side in inequality \eqref{eq:maa} is
 \begin{eqnarray*}
   \sum \limits_{i=1}^{t} \left \lfloor \frac{2\b_i+j+1}{m}  \right\rfloor
&=&\sum_{i=1}^t\left\lfloor\frac{r_i+j}{m}\right\rfloor+\sum_{i=1}^tq_i
 \end{eqnarray*}
 Hence, inequality \eqref{eq:maa} is equivalent to
 \begin{eqnarray*}
   \frac{r(j+1)+\sum_{i=1}^tr_i}{m}-r-\sum_{i=1}^t\left\lfloor\frac{r_i+j}{m}\right\rfloor&\geq& 0
 \end{eqnarray*}
 \begin{eqnarray}
 \label{eq:equivstatement}
    \frac{r(j+1)}{m}-r+\sum_{i=1}^t\left(\frac{r_i}{m}-\left\lfloor\frac{r_i+j}{m}\right\rfloor\right)&\geq& 0
 \end{eqnarray}

 Suppose first that $t<r$ and $r_i=0$ for $i=1,\ldots, t$.
 By $\eqref{eq:jequiv}$ we deduce that $j=m-1$
 and inequality \eqref{eq:equivstatement} easily follows.
 Now suppose $t=r$ and $r_i=r_j$ for every $i,j$.
 By equation \eqref{eq:jequiv} we have 
 $j\equiv -1 -\tilde r r r_1\equiv -1-r_1 \mod m$ and so
    $j=m-1-r_1$. Now, the left part of inequality 
 \eqref{eq:equivstatement} is
     $\frac{r(m-r_1)}{m}-r+r\left(\frac{r_1}{m}-\left\lfloor\frac{m-1}{m}\right\rfloor\right)=0$ and so we have the isometry-dual property.
 
 Conversely, suppose that $t\leq r$ and that the flag is isometry-dual.
Notice that from inequality \eqref{eq:equivstatement} we deduce
$$\left(\frac{(r-t)(j+1)}{m}-r\right)+\left(\sum_{i=1}^t\left(\frac{r_i+j+1}{m}-\left\lfloor\frac{r_i+j}{m}\right\rfloor\right)\right)\geq 0.$$
Now, from the inequality $j\leq m-1$ we know that 
\begin{equation*}
  \left\{
  \begin{array}{l}
    \frac{(r-t)(j+1)}{m}-r\leq -t\\
    \sum_{i=1}^t\left(\frac{r_i+j+1}{m}-\left\lfloor\frac{r_i+j}{m}\right\rfloor\right)\leq t
  \end{array}
  \right.
\end{equation*}
Hence, necessarily
\begin{equation*}
  \left
  \{
  \begin{array}{l}
    \frac{(r-t)(j+1)}{m}-r= -t\\
    \sum_{i=1}^t\left(\frac{r_i+j+1}{m}-\left\lfloor\frac{r_i+j}{m}\right\rfloor\right)=t
  \end{array}\right.
\end{equation*}
That is, either
\begin{center}
$
  \left
  \{
  \begin{array}{l}
    t<r, \quad j=m-1\\
    \frac{r_i}{m}-\left\lfloor\frac{r_i+m-1}{m}\right\rfloor=0 \mbox{ for all }i
  \end{array}\right.
  $
  or \quad
  $
    \left
  \{
  \begin{array}{l}
    t=r\\
    \frac{r_i+j+1}{m}-\left\lfloor\frac{r_i+j}{m}\right\rfloor=1  \mbox{ for all }i
  \end{array}\right.
  $
\end{center}
and so either
\begin{center}
$
  \left
  \{
  \begin{array}{l}
  t<r,\\
    j=m-1,\\
    r_i=0 \mbox{ for all }i,
  \end{array}\right. 
  $
  or \quad
  $
    \left
  \{
  \begin{array}{l}
    t=r,\\
    r_i=m-1-j  \mbox{ for all }i.
  \end{array}\right.
  $
 \end{center}
The theorem follows.
\end{proof}

\begin{example}\label{ex:hermi}
Consider the Hermitian function field over ${\mathbb F}_{16}$ defined by the curve ${\mathcal X}$ with equation $Y^5 + X^4 + X = 0$ of genus $g=6$.
In this case $m=5$ and $r=4$ and the curve has $1$ single point $P_\infty$ at infinity and $4$ other totally ramified places which we may call $Q_1, Q_2, Q_3, Q_4$.

If we take $t=4$, we will obtain flags of 5-point codes by varying $a$ in $C_{\mathcal{L}}(D,aP_\infty+\beta_1Q_1+\beta_2Q_2+\beta_3Q_3+\beta_4Q_4)$, where $D$ is the sum of all rational points except for $P,Q_1,Q_2,Q_3,Q_4$.
Since in this case $t=r$, the flags will satisfy the isometry-dual property whenever $2\beta_1$, $2\beta_2,2\beta_3$, $2\beta_4$ are congruent modulo $5$, for instance, for $(\beta_1,\beta_2,\beta_3,\beta_4)=(3, -2, -7, 8)$.

If we distinguish $3$ of the for ramified points, let us say $Q_1,Q_2,Q_3$, then $t=3$ and 
we will obtain flags of 4-point codes by varying $a$ in $C_{\mathcal{L}}(D,aP_\infty+\beta_1Q_1+\beta_2Q_2+\beta_3Q_3)$, where now $D$ is the sum of all rational points except for $P,Q_1,Q_2,Q_3,Q_4$.
Since in this case $t<r$, the flags will satisfy the isometry-dual property whenever $2\beta_1=1$, $2\beta_2=1,2\beta_3=1$ are congruent to zero modulo $5$, for instance, for $(\beta_1,\beta_2,\beta_3)=(2,-3,7)$.

\end{example}

\begin{example}\label{ex:luciane}
Consider the function field $\mathcal{F}$ over ${\mathbb F}_{49}$ defined by the curve ${\mathcal X}$ with equation $Y^5 - X^2 - X-1 = 0$ of genus $g=2$.

In this case $m=5$ and $r=2$ and the curve has $1$ single point $P_\infty$ at infinity and $2$ other totally ramified places which we may call $Q_1, Q_2$.

If we take $t=2$, we will obtain flags of 3-point codes by varying $a$ in $C_{\mathcal{L}}(D,aP_\infty+\beta_1Q_1+\beta_2Q_2)$, where $D$ is the sum of all rational points except for $P,Q_1,Q_2$.
Since in this case $t=r$, the flags will satisfy the isometry-dual property whenever $2\beta_1$, $2\beta_2$ are congruent modulo $5$, for instance, for $(\beta_1,\beta_2)=(6,1)$.
\end{example}

Now we can prove that for Kummer extensions and flags of codes as in Theorem \ref{t:isodualKummer2} the complete determination of the set $\widehat{H}_\bb^*$ actually depends on the analysis of $2g-1$ values. 

\begin{proposition}\label{elementsHbstar}
With the same notation as in Theorem \ref{t:isodualKummer2}, let $\bb=(\b_1, \dots, \b_t) \in \Z^t$ be such that $\sum\limits_{i=1}^t \b_i< (n+2g-1)/2$. Consider the ${\bf x}$-isometry-dual flag of algebraic geometry codes 
\begin{equation*}
\{0\}=C_\mathcal L(D, a_0P+{\bf G_\bb})\subsetneq
C_\mathcal L(D, a_1P+{\bf G_\bb})\subsetneq \dots \subsetneq C_\mathcal L(D, a_nP+{\bf G_\bb})=\mathbb{F}_{q}^n,
\end{equation*} 
  where $a_0=a_1-1$ and $\{a_1, \dots, a_n\}$ is the ordered set of elements in $\widehat{H}_{\bf \bb}^*$.
Then for $B=\sum\limits_{i=1}^t \b_i$ and
   $$A=\left\{-B\leq a \leq -B+2g-1:
   \text{ the residue }j_0 \text{ of } a\tilde r\mod m \text{ satisfies } 
\frac{a-rj_0}{m} +\sum \limits_{i=1}^{t} \left \lfloor \frac{\b_i+j_0}{m}  \right\rfloor \geq 0\right\},$$
 we have
\begin{align*}
\widehat{H}_\bb^*=A \cup \{2g-B, 2g-B+1, \dots, n-1-B\}\cup \{n+2g-1-2B-a\, :\, a \in A\}.
\end{align*}
In particular, if $\widehat{H}_\bb^*=\{a_1<\cdots <a_n\}$ and $a_0=a_1-1$, then 
the $a_i$'s satisfy the following symmetry condition  $$a_i+a_{n-i}=n+2g-1-2\sum_{i=0}^n \b_i, \quad 0\leq i\leq n. $$ 
\end{proposition}
\begin{proof}
Let $a \in \widehat{H}_{\bf \bb}^*$ and $a^\bot=n+2g-2-2(\b_1+\cdots+\b_t)-a$.
From Lemma \ref{computationHbetastar} we have $\widehat{H}_\bb^*=\{a\in \widehat{H}_\bb: a^\bot+1\in \widehat{H}_\bb\}$.
Applying Corollary \ref{ainHbeta} for $a$ and $a^\bot$ yields: $ a \in \widehat H_\b$ if and only if
\begin{enumerate}[i)]
\item $ a \geq n-\sum\limits_{i=1}^t \b_i$ or 
\item $-\sum\limits_{i=1}^t \b_i \leq a < n-\sum\limits_{i=1}^t \b_i$ and 
 $\exists\,\, 0 \leq j_0 \leq m-1 \text{ such that } m|(a-rj_0)$ and $\frac{a-rj_0}{m} +\sum \limits_{i=1}^{t} \left \lfloor \frac{\b_i+j_0}{m}  \right\rfloor \geq 0$,
\end{enumerate}
and, $ a^\bot+1\in \widehat H_\b$ if and only if
\begin{enumerate}[i)]
\item $a \leq 2g-1-\sum\limits_{i=1}^t \b_i$ or 
\item $2g-1-\sum\limits_{i=1}^t \b_i < a \leq  n+2g-1-\sum\limits_{i=1}^t \b_i$ and 
 $\exists\,\, 0 \leq j_0 \leq m-1 \text{ such that } m|(n+2g-2-2(\b_1+\cdots+\b_t)-a+ 1-rj_0)$ and $ \frac{n+2g-2-2(\b_1+\cdots+\b_t)-a+ 1-rj_0}{m} +\sum \limits_{i=1}^{t} \left \lfloor \frac{\b_i+j_0}{m}  \right\rfloor \geq 0$.
\end{enumerate}

Since $$\left\{2g-\sum_{i=1}^t \b_i, 2g-\sum_{i=1}^t \b_i+1, \dots, n-1-\sum_{i=1}^t \b_i\right\} \subset \widehat{H}_{\bf \bb}^*,$$ 
we are left to analyze the following cases: 
\begin{enumerate}[i)]
\item $-\sum\limits_{i=1}^t \b_i \leq a \leq 2g-1-\sum\limits_{i=1}^t \b_i$ and 
 $\exists\,\, 0 \leq j_0 \leq m-1 \text{ such that } m|(a-rj_0)$ and $ \frac{a-rj_0}{m} +\sum \limits_{i=1}^{t} \left \lfloor \frac{\b_i+j_0}{m}  \right\rfloor \geq 0$, and
\item $n-\sum\limits_{i=1}^t \b_i \leq a \leq n+2g-1-\sum\limits_{i=1}^t \b_i$ and 
 $\exists\,\, 0 \leq j_0 \leq m-1 \text{ such that } m|(n+2g-2-2(\b_1+\cdots+\b_t)-a+ 1-rj_0)$ and $\frac{n+2g-2-2(\b_1+\cdots+\b_t)-a+ 1-rj_0}{m} +\sum \limits_{i=1}^{t} \left \lfloor \frac{\b_i+j_0}{m}  \right\rfloor \geq 0$ .
\end{enumerate}
And the proposition follows. 
 \end{proof}

\begin{example}\label{examples}
To illustrate Proposition \ref{elementsHbstar}, take the same flags of codes as in Example \ref{ex:hermi}.
For $t=4=r$ and $\bb=(3, -2, -7, 8)$, we have $A=\{-2, 2, 3, 6, 7, 8\}$ and therefore it holds
$$\{a_0\}\cup\widehat{H}_\bb^*=\{-3 \} \cup A\cup\{10,\ldots,57\}\cup\{59,60,61,64,65,69\}\;.$$ 

For $t=3<r$ and $\bb=(2, -3, 7)$,  we have  $A=\{-3, 0, 1, 2, 4, 5\}$ and
$$\{a_0\}\cup\widehat{H}_\bb^*=\{-4\}\cup A\cup\{6,\ldots,54\}\cup\{55,56,58,59,60,63\}\;.$$ 

%
%
%
%

\end{example}

\begin{example}
Take now the same flags of codes as in Example \ref{ex:luciane} with $t=2=r$ and $\bb=(6,1)$. In this case we have $A=\{-7,-5\}$ and therefore it holds
$$\{a_0\}\cup\widehat{H}_\bb^*=\{-8 \} \cup A\cup\{-3,\ldots,39\}\cup\{41,43\}\;.$$ 
 \end{example}

\begin{theorem}\label{t:isodualKummer3}
  With the same notation as in Theorem \ref{t:kummerseminfinito}, where $P\neq P_\infty$, let $ 2\leq t \leq r$ and let  $\bb=(\b_2, \dots, \b_t) \in \Z^{t-1}$ be such that $\sum_{i=2}^t \b_i< (n+2g-1)/2$. There exists no constant vector $\bf x$ such that the flag of codes
\begin{equation*}
\{0\}=C_\mathcal L(D, a_0P+{\bf G_\bb})\subsetneq
C_\mathcal L(D, a_1P+{\bf G_\bb})\subsetneq
C_\mathcal L(D, a_2P+{\bf G_\bb})\subsetneq
\dots \subsetneq C_\mathcal L(D, a_nP+{\bf G_\bb})=\mathbb{F}_{q}^n
\end{equation*} 
is $\bf x$-isometry-dual, where $a_0=a_1-1$ and
$a_1<\dots < a_{n}\leq n+2g-1-(\b_2+\cdots+
\b_t)$ is the ordered set of elements in $\widehat{H}_{\bf \bb}^*$.

\end{theorem}
\begin{proof}
By Theorem~\ref{t:isodual}, the flag is ${\bf x}$-isometry-dual for some vector ${\bf x}$ if and only if
  $c:=n+2g-1-2\sum_{i=2}^t \b_i\in \widehat{H}_{2\bb}^*$.
 And by Theorem~\ref{t:kummerseminfinito},
  $c \in \widehat{H}_{2\b}^*$ if and only if  
 \begin{equation}\label{satisfied}
\frac{c+j+1}{m}+\sum\limits_{i=2}^{t} \left \lfloor \frac{2\b_i+j+1}{m}  \right\rfloor -\left \lfloor \frac{n+t+rj}{m}  \right\rfloor <0,
 \end{equation}
where $j$ is the residue of $-c-1$ modulo $m$. For $i=1,\dots , t$ write $2\beta_i+1=mq_i+r_i, q_i \in \Z, 0\leq r_i<m$, then
from $n=r-t+(r_1-r)m$, we get 
\begin{align*} 
c+j+1  & = n+m(r-1)-r-\sum_{i=2}^t 2\b_i +j+1 \\
&= n+m(r-1)-(r-t)+j-\sum_{i=2}^t (mq_i+r_i)\\
&= m(r_1-1)+j-\sum_{i=2}^t (mq_i+r_i).
\end{align*}
So $c+ j+1 \equiv j-\sum_{i=2}^t r_i \mod m$, and, in particular,
  $j\equiv \sum_{i=2}^t r_i \mod m$. 
  
Write $\sum \limits_{i=2}^t r_i-j=mk, 0\leq k\leq t-1$. Then
\begin{align*} 
&\frac{c+j+1}{m}+\sum \limits_{i=2}^{t} \left \lfloor \frac{2\b_i+j+1}{m}  \right\rfloor  - \left \lfloor \frac{n+t+rj}{m}  \right\rfloor\\
&=r-1 -\frac{\sum \limits_{i=2}^{t } r_i-j}{m}  + \sum \limits_{i=2}^{t} \left \lfloor \frac{r_i+j}{m}  \right\rfloor  - \left \lfloor \frac{r(j+1)}{m}  \right\rfloor\\
 &=r-1-k+\sum \limits_{i=2}^{t }\left \lfloor \frac{r_i+\sum \limits_{i=2}^t r_i-mk}{m}  \right\rfloor- \left \lfloor \frac{r(\sum \limits_{i=2}^t r_i-mk+1)}{m}  \right\rfloor\\
  &=r-1-k-(t-1)k+\sum \limits_{i=2}^{t }\left \lfloor \frac{r_i+\sum \limits_{i=2}^t r_i}{m}  \right\rfloor+rk- \left \lfloor \frac{r(\sum \limits_{i=2}^t r_i+1)}{m}  \right\rfloor\\
 & \geq r+k+(r-t)k-1-k+\sum \limits_{i=2}^{t }\left \lfloor \frac{r_i+\sum \limits_{i=2}^t r_i}{m}  \right\rfloor- \left \lfloor \frac{r((t-1)(m-1)+1)}{m}  \right\rfloor\\
  &\geq r-1+(r-t)k- \left \lfloor \frac{r((t-1)(m-1)+1)}{m}  \right\rfloor \\
  &\geq
 r-1-\frac{r(t-1)}{m}-\left \lfloor \frac{-r(t-2)}{m}  \right\rfloor\\
 & \geq
  r-1-\frac{r(t-1)}{m} \geq 0.
  \end{align*}
\end{proof}

\section{Conclusion}

\newcommand\graphh{}
\newcommand\graphhexta{}
\renewcommand\graphh{
\begin{figure}[!htbp]
\caption{ Hermitian  function field over ${\mathbb F}_{ 16 }$ defined by the equation $ Y^5 + X^4Z + XZ^4 =0$. 
Analysis of flags $S^+_b$ satisfying the isometry-dual property.
}
\setlength{\unitlength}{1cm}\newcommand\marcaisodual{{\thicklines\circle{1.01}\circle{1.}\circle{.99}}}
\label{fig:Hermiteisometrydualfour}
\begin{center}
\begin{minipage}{ 0.900000000000000 \textwidth}
\resizebox{\textwidth}{!}{

at $(a,b)$ if $C_{{\mathcal L}}(D,aP+bQ)$
in an isometry-dual flag guaranteed 
 in \cite{BCQ2021}
}\end{minipage}\end{minipage}
\end{center}\end{figure}
}
\renewcommand\graphhexta{
\begin{figure}[!htbp]
\caption{ Hermitian  function field over ${\mathbb F}_{ 16 }$ defined by the equation $ Y^5 + X^4Z + XZ^4 =0$. 
Analysis of flags $S_b$ satisfying the isometry-dual property.
}
\setlength{\unitlength}{1cm}\newcommand\marcaisodual{{\thicklines\circle{1.01}\circle{1.}\circle{.99}}}
\label{fig:Hermiteisometrydualextafour}
\begin{center}
\begin{minipage}{ 0.900000000000000 \textwidth}
\resizebox{\textwidth}{!}{

at $(a,b)$ if $C_{{\mathcal L}}(D,aP+bQ)$
in an isometry-dual flag guaranteed 
by Theorem~\ref{t:isodual}.
}\end{minipage}\end{minipage}
\end{center}\end{figure}
}

\paragraph{Previous work.}
In \cite{BCQ2021} we
analyzed flags of two-point codes of the form $C_{\mathcal L}(D,aP+bQ)$.
Let $H_b^*=\{a\in\Z \, :\, C_\mathcal L(D, aP+bQ)\neq C_\mathcal L(D, (a-1)P+bQ) \}$. More precisely, in that reference we studied the flags of codes
$$S_b^+:=\left(C_{\mathcal L}(D,aP+bQ)\right)_{a\in \{0\}\cup(H_b^*\cap\{1,\dots,n+2g-2-2b\})}$$
and we obtained a characterization of the non-negative integers $b$ in the range $0\leq b\leq n/2-g-1$ for which $S_b^+$ satisfies the isometry-dual property.

For instance, for the case of 
the Hermitian function field over ${\mathbb F}_{16}$ defined by the curve ${\mathcal X}$ with equation
$Y^5 + X^4Z + XZ^4 = 0$ of genus $g=6$, taking $P = (1 : 0 : 0)$ the place at infinity, $Q = (0 : 0 : 1)$, and $D$ the sum of all rational places of ${\mathcal X}$ except $P$ and $Q$ (that is, $n=63$),
the set of values of $b$ in the aforementioned range for which
$S_b^+$ satisfies the isometry-dual property are exactly the values in the set $\{2,7,12,17,22\}$.
The corresponding flags of codes $S_2^+$, $S_7^+$, $S_{12}^+$, $S_{17}^+$, $S_{22}^+$ are remarked with circles in Figure~\ref{fig:Hermiteisometrydualfour}, where a bullet at position $(a,b)$ refers to the code $C_{\mathcal L}(D,aP+bQ)$ and flags correspond to the bullets in a horizontal line.

\graphh

\paragraph{From non-negative indices to any index.}
The results we present now in the present paper allow to extend the results in \cite{BCQ2021} to the full flags of codes
$$S_b:=\left(C_{\mathcal L}(D,aP+bQ)\right)_{a\in H_b^*\cup\{-1+\min H_b^*\}},$$
where now $a$ ranges in $H_b^*\cup\{-1+\min H_b^*\}$ instead of $\{0\}\cup(H_b^*\cap\{1,\dots,n+2g-2-2b\})$.
Hence, we can consider the full flags of codes depicted in Figure~\ref{fig:Hermiteisometrydualextafour}, where $a$ can be either positive or negative.

\paragraph{From fixed coefficients restricted to a range to any set of fixed coefficients.}
Also, while in \cite{BCQ2021} the results were restricted to the range $0\leq b\leq n/2-g-1$ ($0\leq b\leq 24$ in the example), now we prove the analogous results for any integer $b$. In Figure~\ref{fig:Hermiteisometrydualextafour} we depicted all the flags $S_b$ satisfying the isometry-dual property in the range $-25\leq b\leq 25$ for the sake of readability, but we could have chosen any range. The flags satisfying the property are exactly $\dots,S_{-23},S_{-18},S_{-13},S_{-8},S_{-3}, S_2$, $S_7$, $S_{12}$, $S_{17}$, $S_{22},\dots$.

\graphhexta

\paragraph{From two point codes to many-point codes.}
Another improvement is that instead of fixing one single place $Q$ and an integer $b$, now we fix a set of places $Q_1,\dots,Q_t$ and a divisor ${\bf G_\beta}=\beta_1Q_1+\dots+\beta_tQ_t$.
For this reason we need to introduce the set $\widehat{H}_{(\beta_1,\dots,\beta_t)}^*=\{a\in\Z \, :\, C_\mathcal L(D, aP+{\bf G_\beta})\neq C_\mathcal L(D, (a-1)P+{\bf G_\beta}) \}$. Notice that for $t=1$,
$\widehat{H}_{\beta_1}^*$ coincides with $H_{\beta_1}^*$.
  Then we
analyze the full flags 
$$S_{(\beta_1,\dots,\beta_t)}:=\left(C_{\mathcal L}(D,aP+\beta_1Q_1+\dots+\beta_tQ_t)\right)_{a\in \widehat H_{(\beta_1,\dots,\beta_t)}^*\cup\{-1+\min \widehat H_{(\beta_1,\dots,\beta_t)}^*\}}$$
In this case we also obtain characterizations of the vectors $(\beta_1,\dots,\beta_t)$ for which the flag $S_{(\beta_1,\dots,\beta_t)}$ satisfies the isometry-dual property.
See, for instance, the examples already presented in Figure~\ref{fig:QQDKleinisometryduallarge} and Figure~\ref{fig:QQGKleinisometryduallarge}.

\section{Acknowledgment}
The authors would like to thank the anonymous referees for comments that improved the results and the presentation of this work. The authors would like to thank the funding from FAPERJ
\#88881.360642/2019-01 that made their collaboration possible. They
would also like to thank Kwankyu Lee for his help in the usage of {\sc SAGE}.
The first author was partially supported by Consell Català de Recerca i
Innovació (2021 SGR 00115) and Ministerio de Ciencia e Innovación (ACITHEC PID2021-124928NB-I00). The second
author was partially supported by Fundação de Amparo à Pesquisa do Estado
de Minas Gerais (FAPEMIG APQ-00696-18 and RED-00133-21). The third author would like to thank CNPq 302727/2019-1 and FAPERJ 260003/002364/2021.

\bibliographystyle{plain}

\end{document}